\RequirePackage{fix-cm}
\documentclass{svjour3}
\usepackage{graphicx}
\usepackage{epstopdf}
\usepackage{amsmath}
\usepackage{amsfonts}
\usepackage{amssymb}
\usepackage{bm}
\usepackage{array,multirow,mathabx,longtable}
\usepackage{subfigure}
\usepackage[symbol]{footmisc}

\newcommand{\nn}{{\nonumber}}
\newcommand{\ba}{\begin{eqnarray}}
\newcommand{\ea}{\end{eqnarray}}

\begin{document}

\title{Few-body systems consisting of mesons.}
\author{A. Mart\'inez Torres \and K. P. Khemchandani \and L. Roca \and E. Oset}

\institute{Universidade de Sao Paulo, Instituto de Fisica, C.P. 05389-970, Sao 
Paulo, Brazil.\\ \email{amartine@if.usp.br} \and Universidade Federal de Sao Paulo, C.P. 01302-907, Sao Paulo, Brazil. \email{kanchan.khemchandani@unifesp.br} \and Departamento de F\`isica, Universidad de Murcia, E-30071 Murcia, Spain.\email{luisroca@um.es} \and Departamento de F\'isica Te\'orica e IFIC, Centro Mixto Universidad de Valencia-CSIC, Institutos de Investigaci\'on de Paterna, Apdo 22085, 46071 Valencia, Spain. \email{oset@ific.uv.es}}

\date{\today}
\maketitle
\begin{abstract}
\noindent
We present a work which is meant to inspire the few-body practitioners to venture into the study of new, more exotic, systems and to hadron physicists, working mostly on two-body problems, to move in the direction of studying related few-body systems. For this purpose we devote the discussions in the introduction to show how the input two-body amplitudes can be easily obtained using techniques of the chiral unitary theory, or its extensions to the heavy quark sector. We then briefly explain how these amplitudes can be used to solve the Faddeev equations or a simpler version obtained by treating the three-body scattering as that of a particle on a fixed center. Further, we give some examples of the results obtained by studying systems involving mesons. We have also addressed the field of many meson systems, which is currently almost unexplored, but for which we envisage a bright future. Finally, we give a complete list of works dealing with unconventional few-body systems involving one or several mesons, summarizing in this way the findings on the topic, and providing a motivation for those willing to investigate such systems.
\end{abstract}


\section{Introduction}
   This paper is not supposed to be a review on nonconventional few-body systems made out of other hadrons than the baryons normally studied as the bulk of few-body physics. It is instead offering a view of new developments in these areas with the purpose to provide tools and a perspective to few-body practitioners of the immense field opened in few-body physics by recent developments in hadron physics. 

      The first thing to realize is that similarly to the forces generated between nucleons, there is also and interaction between mesons, that sometimes is stronger than the known nuclear forces. Why are there not many body systems made only from mesons is then not the question, but are there such systems and we failed to identify them? The novelty with mesons is that, unlike systems with baryons where the baryon conservation number is a stabilizing factor, there is no meson conservation number and systems with many mesons can decay into systems with fewer mesons for which there is always a big phase space for decay. If we pile up many mesons the decay width will increase. The proper question is then: how many mesons can we pile up before the width becomes so large that the object cannot be identified experimentally as a particle?

    Certainly one can also have systems with a meson and a baryon, or a nucleus, which have for long been studied, but newer systems would contain two or more mesons and a baryon, or a mixture of them. If one adds to this the new developments in the charm and bottom sector in hadron physics, the potential for new few-body systems is simply huge, and its study is only in its infancy. The devotion of few-body physicists to such problems should lead in the future to a rich field of relevant physics. 

    Two ingredients have to be discussed for this purpose: 1) How to get the forces between mesons among themselves, and mesons with baryons. 2) Are there some new findings that should be considered in the technical solution of the few-body equations? We devote some words to these issues in what follows. 

\section{Meson-meson interaction} 
  The good news about this sector is that the introduction and popularization of the chiral Lagrangians has brought new and accurate tools to deal with this problem. Originally introduced by Weinberg \cite{weinberg} and systematized later in the SU(3) space \cite{gasser}, the chiral Lagrangians offer an effective theory of QCD at low energies, where the basic symmetries, chiral symmetry and other important ones, are incorporated in Lagrangians where the degrees of freedom are the mesons, not the quarks and gluons. They are ready to extract from them the meson-meson potentials.   At lowest order, sufficient for most purposes if they are properly unitarized (see later), the chiral Lagrangian reads
\begin{align}
\mathcal{L}=\frac{1}{12 f^2}\langle (\partial_\mu \Phi\Phi-\Phi\partial_\mu \Phi)^2+M\Phi^4\rangle, 
\end{align}
where the symbol $\langle\quad\rangle$ indicates the trace in the flavour space of the SU(3) matrices appearing in $\Phi$  and $M$, $f$ is the pion decay constant (93 MeV) and the matrices $\Phi$ and $M$ are given by 
\begin{align}
\Phi=\left(\begin{array}{ccc}\frac{1}{\sqrt{2}}\pi^0+\frac{1}{\sqrt{6}}\eta_8&\pi^+&K^+\\\pi^-&-\frac{1}{\sqrt{2}}\pi^0+\frac{1}{\sqrt{6}}\eta_8&K^0\\
K^-&\bar K^0&-\frac{2}{\sqrt{6}}\eta_8\end{array}\right),\label{Phi}
\end{align}
\begin{align}
M=\left(\begin{array}{ccc}m^2_\pi&0&0\\0&m^2_\pi&0\\0&0&2m^2_K-m^2_\pi\end{array}\right).
\end{align}

   To show that dealing with this is easier than one might think, the tree level amplitudes, let us say the potentials that come from there, are evaluated in Ref.~\cite{npa}, and projected over s-wave for the coupled meson-meson channels in strangeness zero, charge zero and are given by \cite{liangoset}
\begin{align}
V=\left(\begin{tabular}{ccccc}
$-\frac{s}{2f^2}$&$-\frac{1}{\sqrt{2}f^2}(s-m^2_\pi)$&$-\frac{s}{4f^2}$&$-\frac{s}{4f^2}$&$-\frac{m^2_\pi}{3\sqrt{2}f^2}$\\
$V_{12}$&$-\frac{m^2_\pi}{2f^2}$&$-\frac{s}{4\sqrt{2}f^2}$&$-\frac{s}{4\sqrt{2}f^2}$&$-\frac{m^2_\pi}{6f^2}$\\
$V_{13}$&$V_{23}$&$-\frac{s}{2f^2}$&$-\frac{s}{4f^2}$&$-\frac{9s-6m^2_\eta-2m^2_\pi}{12\sqrt{2}f^2}$\\
$V_{14}$&$V_{24}$&$V_{34}$&$-\frac{s}{2f^2}$&$-\frac{9s-6m^2_\eta-2m^2_\pi}{12\sqrt{2}f^2}$\\
$V_{15}$&$V_{25}$&$V_{35}$&$V_{45}$&$-\frac{16m^2_K-7m^2_\pi}{18 f^2}$
\end{tabular}\right),\label{V0}
\end{align}
where the indices 1 to 5 refer to the states 1 for $\pi^+\pi^-$, 2 for $\pi^0\pi^0$, 3 for $K^+ K^-$, 4 for $K^0\bar K^0$ and 5 for $\eta\eta$.  These states build amplitudes in isospin $I=0$. If we want to also account for $I=1$, one has to introduce the $\pi \eta$ channel (index 6) and three more matrix elements \cite{daixie}
\begin{align}
V=\left(\begin{tabular}{c|c}\text{Same as in Eq.~(\ref{V0})}&\begin{tabular}{c}0\\0\\$-\frac{\sqrt{3}}{12f^2}\left(3s-\frac{8}{3}m^2_K-\frac{m^2_\pi}{3}-m^2_\eta\right)$\\
$\frac{\sqrt{3}}{12f^2}\left(3s-\frac{8}{3}m^2_K-\frac{m^2_\pi}{3}-m^2_\eta\right)$\\0\end{tabular}\\
\hline
\begin{tabular}{ccccc}$V_{16}$&$V_{26}$&$V_{36}$&$V_{46}$&$V_{56}$\end{tabular}&$-\frac{m^2_\pi}{3f^2}$\end{tabular}\right).\label{V1}
\end{align}

These functions are written in momentum space and depend only on the center of mass energy of the pair of mesons ($\sqrt{s}$). To make clear the normalization, let us establish from the beginning that with one of these amplitudes we would write the cross section as 
\begin{align}
\frac{d\sigma_{ij}}{d\Omega}=\frac{1}{64 \pi^2 s}\frac{p_j}{p_i}|V_{ij}|^2,
\end{align}
where $p_i$ ($p_j$) is the modulus of the center or mass momenta of the particles in the initial (final) channel. 

   One may wonder where the range of the interaction appears since Eqs.~(\ref{V0}) and (\ref{V1}) only depend on $s$. The range is important when solving the Lippmann Schwinger equation with these potentials and the range is taken into account in the subsequent step to these chiral Lagrangians which is the so called chiral unitary approach. The Lippmann Schwinger equations, although it has become customary to call them the Bethe Salpeter equation (BSE) since the derivation is done relativistically using a $d^4 q$ integration in the loops, are given in matrix form ($N\times N$, with $N$ being the number of channels) as
\begin{align}
 T= [1-VG]^{-1} V,\label{bse}   
\end{align}
which is an algebraic equation, where $G$ is the loop function of the integral of two propagators \cite{npa}, diagonal in the channels, and can be written as 

\begin{align}
G_{ii}(s)=\int\limits_{0}^{q_\text{max}}\frac{dq\, q^2}{(2\pi)^2}\frac{\omega_1+\omega_2}{\omega_1\omega_2[{P^{0}}^2-(\omega_1+\omega_2)^2+i\epsilon]},
\label{eq:Gii}
\end{align}
where $\omega_i = \sqrt{q^2+m_i^2}$, ${P^0}^2 = s$ and the subindex $i$ stands for the two intermediate states of the $i$ channel. In Eq.~\eqref{eq:Gii}, $q_\text{max}$ is a cut off in the modulus of the three momentum, and this is what in practical terms gives the range of the interaction. This cut off is not an element given by the theory and it is obtained by fitting some experimental observable. In \cite{liangoset,daixie} it is taken as 600 MeV to reproduce relevant data on meson-meson scattering.

   One may also wonder how can we obtain an algebraic equation from the coupled channel integral equations. There are many ways to see that. One of them is based on the use of a dispersion relation \cite{ollerulf}, but for the few-body community the most clarifying proof can be found in \cite{danijuan} where it is shown that Eq.~(\ref{bse}) can be obtained from a separable potential 
\begin{align}
V(\vec{q},\vec{q}^\prime)= V \Theta(q_\text{max}-|\vec{q}|) \Theta(q_\text{max}-|\vec{q}^\prime|),
\label{eq:Vcutoff}
\end{align}
which leads to Eq.~(\ref{bse}) with a scattering matrix
\begin{align}
T \Theta(q_\text{max}-|\vec{q}|) \Theta(q_\text{max}-|\vec{q}^\prime|). 
\end{align}

   It is interesting to mention that the chiral unitary approach solving the BSE gives rise to the low energy resonances $f_0(500)$ (or $\sigma$), $f_0(980)$, $a_0(980)$, $K^*_0(700)$ (or $\kappa$) \cite{npa,kaiser,locher,juanenri}, by using just one cut off to regularize the loops of the order of 600 MeV. In some cases, when a channel is dominant, one can say that these states are kind of molecules of these channels, decaying into other open channels. For instance, the $f_0(980)$ would qualify mostly as a $K \bar K$ bound state, something that was already suggested prior to the developments of the chiral unitary approach in \cite{isgur}. 

    The method described here can easily be extended to the interaction of pseudoscalar-vector interaction \cite{lutz,rocasingh,chengeng} and the interaction of vector-vector \cite{raquel,gengvec,raquelgeng,dispersion}.

\section{Meson-baryon interaction}

The interaction of light mesons with baryons  is also given by the chiral Lagrangians  \cite{ecker,bernard} as 
\begin{align}
\mathcal{L}&=\langle \bar Bi\gamma^\mu\nabla_\mu\rangle -M_B\langle \bar B B\rangle+\frac{1}{2}\langle\bar B\gamma^\mu\gamma_5\left(D\{u_\mu,B\}
+F[u_\mu,B]\right)\rangle,\\
&\nabla_\mu B=\partial_\mu B+[\Gamma_\mu,B],\quad \Gamma_\mu=\frac{1}{2}(u^\dagger\partial_\mu u+u\partial_\mu u^\dagger),\\
&U\equiv u^2=e^{i\frac{\sqrt{2}\Phi}{f}},~\quad u_\mu=iu^\dagger \partial_\mu  U u^\dagger,
\end{align}
where $\Phi$ is given by Eq.~(\ref{Phi}) and
\begin{align}
B=\left(\begin{array}{ccc}\frac{1}{\sqrt{2}}\Sigma^0+\frac{1}{\sqrt{6}}\Lambda&\Sigma^+&p\\\Sigma^-&-\frac{1}{\sqrt{2}}\Sigma^0+\frac{1}{\sqrt{6}}\Lambda&n\\
\Xi^-&\bar \Xi^0&-\frac{2}{\sqrt{6}}\Lambda\end{array}\right).\label{B}
\end{align}

At the lowest order for the meson-baryon interaction, the expansion in the fields of the former Lagrangian gives~\cite{ramoskaon}
\begin{align}
\mathcal{L}=\frac{1}{4f^2}\langle \bar Bi\gamma^\mu[[\Phi,\partial_\mu \Phi],B]\rangle.
\end{align}

  As an example we look at the interaction of $\bar K N$ with its coupled channels. Following \cite{ramoskaon} we have the channels $K^-p$, $\bar K^0n$, $\pi^0 \Lambda$, $\pi^0 \Sigma^0$, $\pi^+ \Sigma^-$, $\pi^- \Sigma^+$, $\eta \Lambda$, $\eta \Sigma^0$, $K^+ \Xi^-$, $K^0 \Xi^0$ and the interaction is given by 
\begin{align}
V_{ij}=-\frac{C_{ij}}{4f^2}(k^0+{k^\prime}^0),
\end{align}
where $k^0$, ${k^\prime}^0$ are the energies of the initial and final mesons. The coefficients $C_{ij}$ are numbers of the order of unity and are tabulated in \cite{ramoskaon}. 

    As one can see, the transition potentials are very easy.  As shown in  \cite{ramoskaon,weise} , this interaction, with a cut off of 630 MeV in \cite{ramoskaon} provides and excellent description of the $\bar K N$ transition to different channels, and interestingly, it gives rise to two $\Lambda(1405)$ states 
\cite{ollerulf,cola}.

\section{Extrapolation to the charm and beauty sectors}
     Hadron physics has had its natural expansion in the field of heavy quarks with charm or beauty~\cite{karliner,chen,crede,klempt,bigi,korner,cheng}. It is not risky to guess that few-body physics will also expand into the area and some incursions have already been done. But before we report on some work along this line, let us first address how the interaction of mesons, or mesons with baryons can be obtained in this realm. Given the aversion that many practitioners have on the use of SU(4) symmetry, the comments we make below are most useful. 
    While it is possible to use SU(4) symmetry on the basic Lagrangians and justify a posteriori the results obtained \cite{montana}, it is interesting to see that one does not need to invoke SU(4) symmetry to extend the idea of the chiral Lagrangians to the heavy quark sector. The key to understand this point is the realization that the chiral Lagrangians in SU(3) can be obtained using the local hidden gauge approach \cite{hidden1,hidden4,hideko} exchanging vector mesons between mesons, or between mesons and baryons. A formal derivation for mesons is done in \cite{rafael}.  A practical example to evaluate the $\bar K N$ interaction from this perspective and the equivalence to the results in the former section can be seen in \cite{debasliang}.

\begin{figure}
 \centering
 \includegraphics[width=0.5\textwidth]{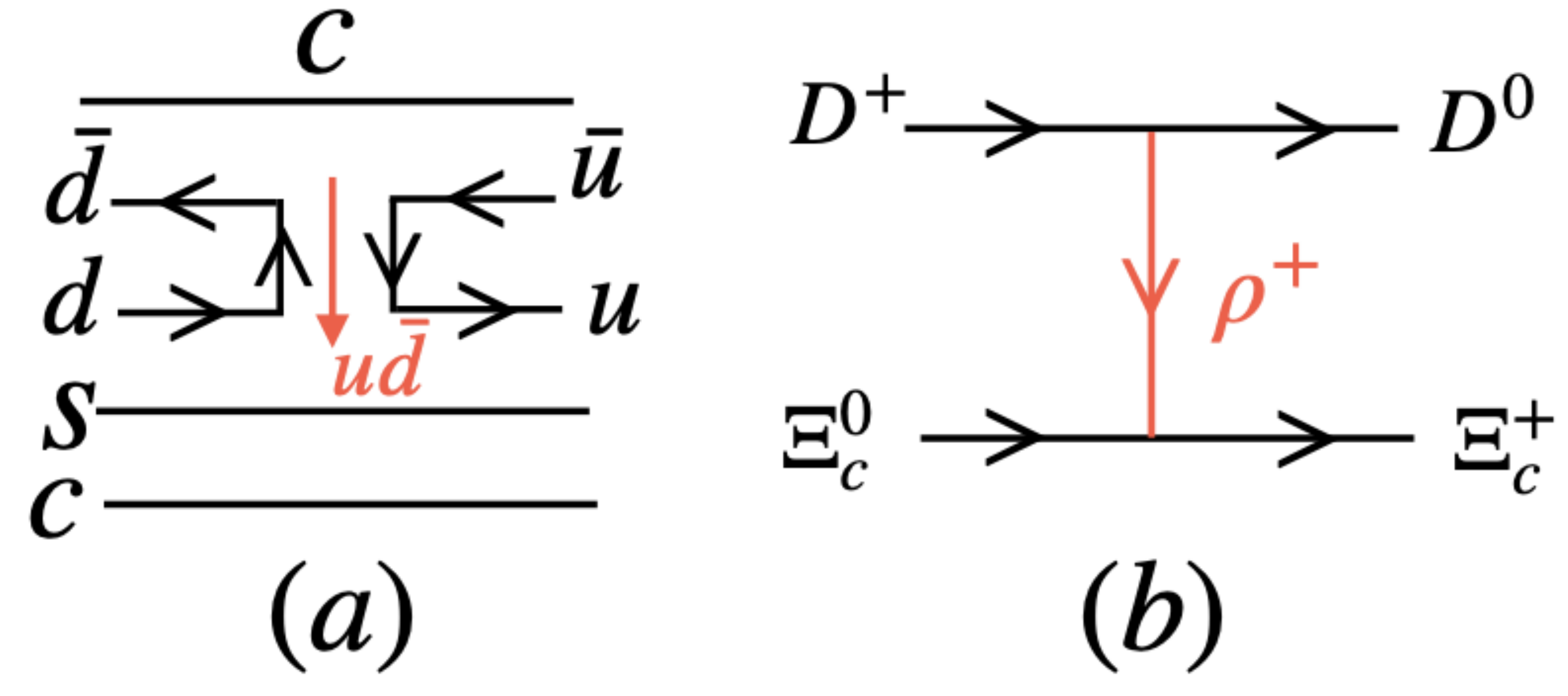}
 \caption{(a) Exchange of a $u\bar d$ meson in the $D^+\Xi^0_c\to D^0 \Xi_c^+$ transition. (b) Macroscopic picture showing $\rho^+$ exchange.}\label{quarks}
 \end{figure}

In the heavy quark sector, when constructing baryons with heavy quarks, one does not write flavor symmetric wave functions in all quarks but instead isolates the heavy quarks and imposes the symmetry on the light quarks. This is a good starting point used in different works \cite{capstick,roberts}. Then let us do an example such that the reader appreciates how easy the evaluation of the interaction is. 
Let us evaluate the transition from  $D^+ \Xi_c^0 \to D_0 \Xi_c^+$. 
\begin{align}
\Xi^+_c=\frac{1}{\sqrt{2}}c(us-su),\quad \Xi^0_c=\frac{1}{\sqrt{2}}c (ds-sd),
\end{align}
and the spin is $\chi_c \chi_\text{MA}(us)=\chi_c\frac{1}{\sqrt{2}}(\updownarrows-\downuparrows)$, with $\chi_c$ being the spin function for the $c$ quark. The exchange of a vector at the quark level is shown in Fig.~\ref{quarks}.
 We have two vertices, the $D^+\rho^+D^0$ and the $\Xi^0_c\rho^+\Xi^+_c$. The first vertex is the ordinary $VPP$ ($V\equiv$ Vector, $P\equiv$ Pseudoscalar) given by the Lagrangian
\begin{align}
\mathcal{L}_{VPP}=-ig\langle [P,\partial_\mu P]V^\mu\rangle,\label{VPP}
\end{align}
with $g=\frac{M_V}{2f}$ ($M_V\simeq 800$ MeV, $f=f_\pi=93$ MeV, the pion decay constant), and $P$, $V$ given by~\cite{debasliang}
\begin{align}
P&=\left(\begin{array}{cccc}\frac{1}{\sqrt{2}}\pi^0+\frac{1}{\sqrt{3}}\eta+\frac{1}{\sqrt{6}}\eta^\prime&\pi^+&K^+&\bar D^0\\\pi^-&-\frac{1}{\sqrt{2}}\pi^0+\frac{1}{\sqrt{3}}\eta+\frac{1}{\sqrt{6}}\eta^\prime&K^0&D^-\\
K^-&\bar K^0&-\frac{1}{\sqrt{3}}\eta+\sqrt{\frac{2}{3}}\eta^\prime&D^-_s\\D^0&D^+&D^+_s&\eta_c\end{array}\right),\label{P}
\end{align}
\begin{align}
V&=\left(\begin{array}{cccc}\frac{1}{\sqrt{2}}\rho^0+\frac{1}{\sqrt{2}}\omega&\rho^+&K^{*+}&\bar D^{*0}\\\rho^-&-\frac{1}{\sqrt{2}}\rho^0+\frac{1}{\sqrt{2}}\omega&K^{*0}&D^{*-}\\
K^{*-}&\bar K^{*0}&\phi&D^{*-}_s\\D^{*0}&D^{*+}&D^{*+}_s&J/\psi\end{array}\right),\label{V}
\end{align}
where $P$ includes the $\eta-\eta^\prime$ mixing~\cite{bramon}.

It might look like by writing Eqs.~(\ref{P}) and (\ref{V}) one is using SU(4) but this is not the case since both matrices are simply the matrices of $q\bar q$ written in terms of the mesons. The use of Eq.~(\ref{VPP}) simply counts the number of quarks and gives the vertex the required structure~\cite{sakairoca}.

The Lagrangian of Eq.~(\ref{VPP}) for $D^+\to \rho^+ D^0$ gives
\begin{align}
-i\mathcal{L}=-ig\rho^{-\mu}(\bar D^0\partial_\mu D^+-D^+\partial_\mu \bar D^0),
\end{align}
and an amplitude for the $D^+\rho^+D^0$ vertex
\begin{align}
-it=i\mathcal{L}=g\epsilon^\mu_{\rho^-} (-i p_{D^+}-ip_{D^0})_\mu.\label{vector}
\end{align}
The Lagrangian for the lower vertex is given by 
\begin{align}
\mathcal{L}_{BVB}=\langle B_i|g\rho^+_\mu\gamma^\mu|B_i\rangle,\label{BVB}
\end{align}
with $B_i$ ($B_f$) being the initial (final) baryons and $\gamma^\mu$ the Dirac matrix, acting on the $B_i$ ($B_f$) spinors. In few-body problems, we will look for stable or resonant states with small three-momenta compared to the heavy masses that we are dealing with. Hence, both in Eq.~(\ref{vector}) and in Eq.~(\ref{BVB}) the zero component dominates and $\gamma^\mu\to \gamma^0\simeq 1$ acting on Pauli spinors. Then, Eq.~(\ref{BVB}) becomes
\begin{align}
\mathcal{L}_{BVB}=\Big\langle c\frac{1}{\sqrt{2}}(us-su)\Big|gu\bar d\Big|c\frac{1}{\sqrt{2}}(ds-sd)\Big\rangle=\frac{1}{2}(2g)=g,
\end{align}
with the $\rho$ propagator
\begin{align}
-\frac{g^{00}}{{q^0}^2-\vec{q}^2-m^2_\rho}\simeq\frac{1}{m^2_\rho},
\end{align}
the transition potential becomes
\begin{align}
V=-\frac{g^2}{m^2_\rho}(p^0_{D^+}+p^0_{D^0})=-\frac{1}{4f^2}(p^0_{D^+}+p^0_{D^0}).
\end{align}

With this exercise the reader can see that the evaluation of the interaction is actually easy and that all matrix elements are of the type of $p^0+{p^\prime}^0$ as in the case of the former section. Examples of these systems solved along these lines can be found in \cite{debasliang,liangdias,diasdebas,yupavao,yudias,diasyu}.

 One can see that we have kept the $c$ quark as a spectator and it has not played any role in the interaction. This is most welcome since automatically these amplitudes will fulfill the rules of heavy quark symmetry \cite{wise,neubert,manohar} which offer an alternative method to evaluate matrix elements with heavy quarks. One should note, however, that one could also exchange a $c\bar c$ object in Fig. 1(a). However, this is highly suppressed because of the heavy mass in the $J/\psi$ propagator and this term is negligible. In other cases one can exchange $D^*$ which is also suppressed, but one can keep it in the calculations. The message is that the dominant terms come from the exchange of light vectors, then the heavy quarks are spectators and heavy quark symmetry rules are automatically fulfilled.

\section{Three-body systems}
\begin{figure}
\centering
\includegraphics[width=0.6\textwidth]{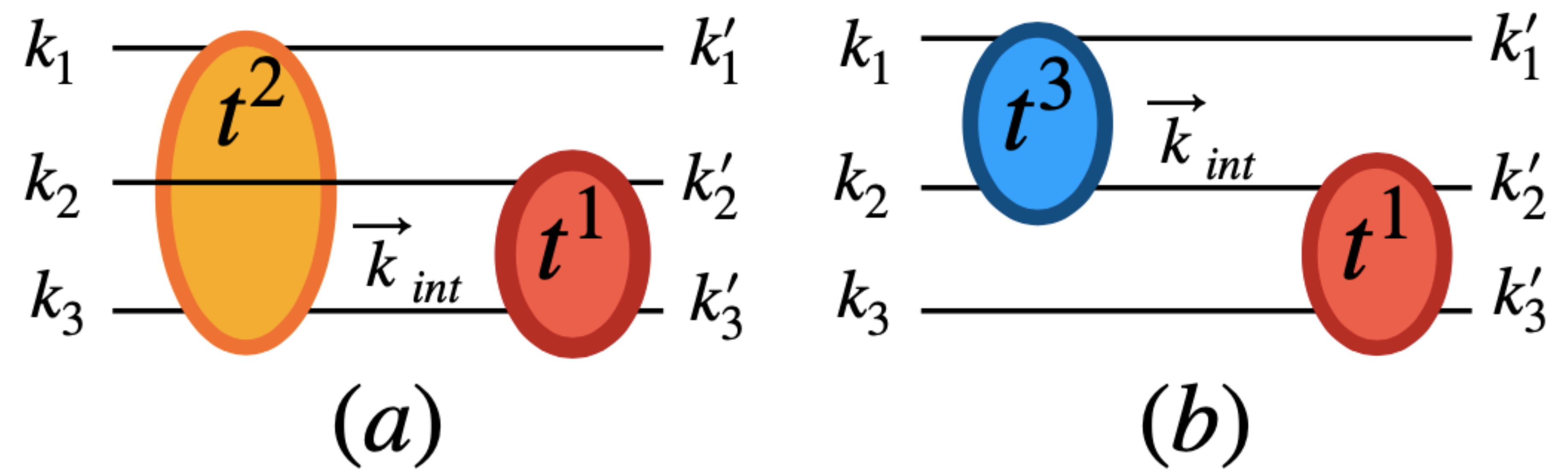}
\caption{Diagrammatic representation of the terms (a) $t^1 g^{12}t^2$ and (b) $t^1 g^{13}t^3$.}\label{tgt}
\end{figure}

     The former sections have shown how to evaluate the two-body potentials, then the $T$ matrices by means of the Bethe-Salpeter equations and then one has all the ingredients to use for instance Faddeev equations, which are the most common tools to solve the three-body equations. The input in these equations are precisely the $T$ matrices, which we have learned to evaluate in the chiral unitary approach, or its extension to the heavy sector by means of the use of the local hidden gauge approach. Certainly one is welcome to use their preferred potentials and $T$ matrices, but the use of the amplitudes that we have discussed have a non trivial consequence that we want to show here. Indeed, Faddeev equations sum diagrams as shown in Fig.~\ref{tgt}.
 
\begin{figure}
\centering
\includegraphics[width=0.55\textwidth]{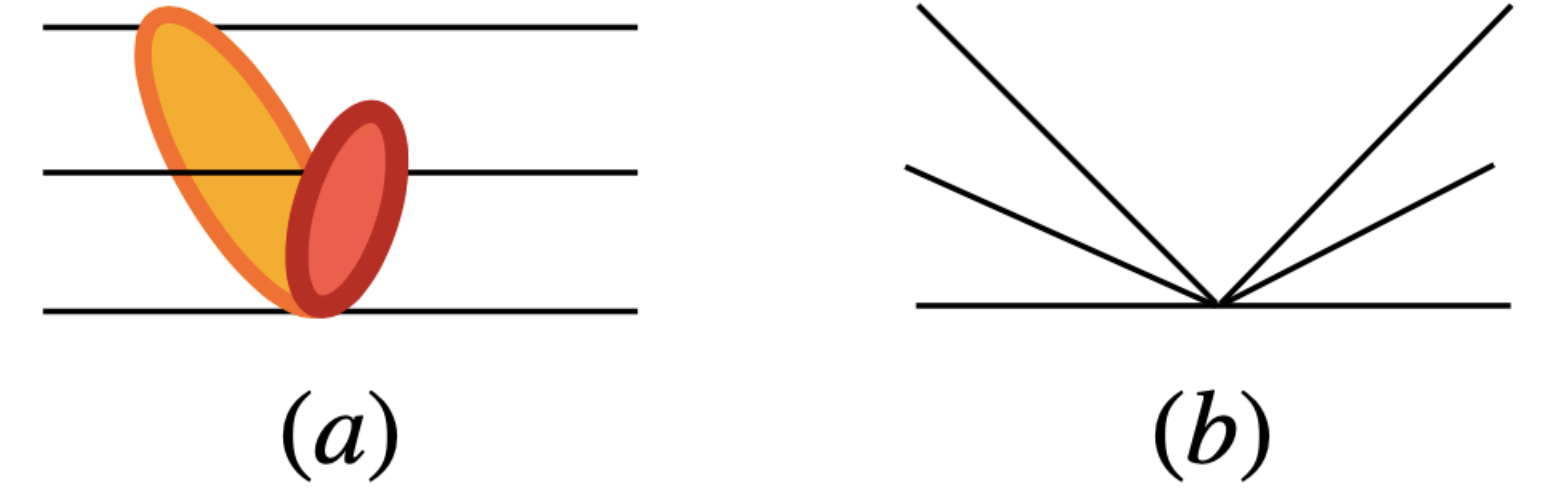}
\caption{The origin of the three-body forces $(a)$ due to cancellation of the propagators in Fig.~\ref{tgt}(a) with the off-shell part of the chiral amplitudes $(b)$ at the tree level from the chiral Lagrangian.}\label{contact}
\end{figure}

   Let us take diagram (a). Imagine that the line $k_\text{int}$ corresponds to a meson. 
We can split the amplitude into a term that we call $t_\text{on}$, which corresponds to taking $k_\text{int}^2 = m^2$ in the amplitude, and the rest, that we call $t_\text{off}$. One can prove that $t_\text{off}$ is proportional to $k_\text{int}^2 - m^2$ ( indeed, by definition it has to be zero for $k_\text{int}^2 = m^2$).  Then $t_\text{off}$ cancels the propagator of this meson in the amplitude and gives rise to a contact term as shown in Fig.~\ref{contact}(a).

  The novelty about using chiral Lagrangians is that upon expansion of the Lagrangians in the fields to have the amplitude for 3 particles $\to$ 3 particles one also gets a contact term (see Fig.~\ref{contact} (b)) and the interesting thing is that the contributions of diagrams (a) and (b) of Fig.~\ref{contact} cancel. This can be seen in detail in \cite{phi2175}. For the few-body practitioner it is not necessary to go trough the detail of the derivation. It suffices to know that the input needed is only the on-shell amplitudes and that the off-shell part does not play any role in the Faddeev equations. This feature is certainly most welcome, because it is well known that the final results cannot depend on off-shell amplitudes since off-shell amplitudes can be changed arbitrarily by means of unitary transformations that, however, do not change the on-shell part, and they are, thus, unphysical. The message is then clear and can be extrapolated for other schemes not using explicitly chiral amplitudes: Use only the on-shell amplitudes in the Faddeev approach. 

      While this is possible in schemes based on the Faddeev equations that has as an elementary input  two particle amplitudes, it is less clear how to implement it when one does for instance variational calculations based on potentials and trial wave functions. Since there are many potentials that generate the same on-shell amplitudes but different off-shell ones, it is then inevitable that the results will depend on the potential used. In most cases one  resorts to taking explicit three-body forces, but two things are clear: first, these three-body forces are not absolute, they depend on the two-body potential used, and second, one of the roles of these three-body forces is to cancel the effect of the off-shell two-body amplitudes that we have discussed. 

\section{Chiral Lagrangians and the Faddeev equations}
\begin{figure}
\centering
\includegraphics[width=0.55\textwidth]{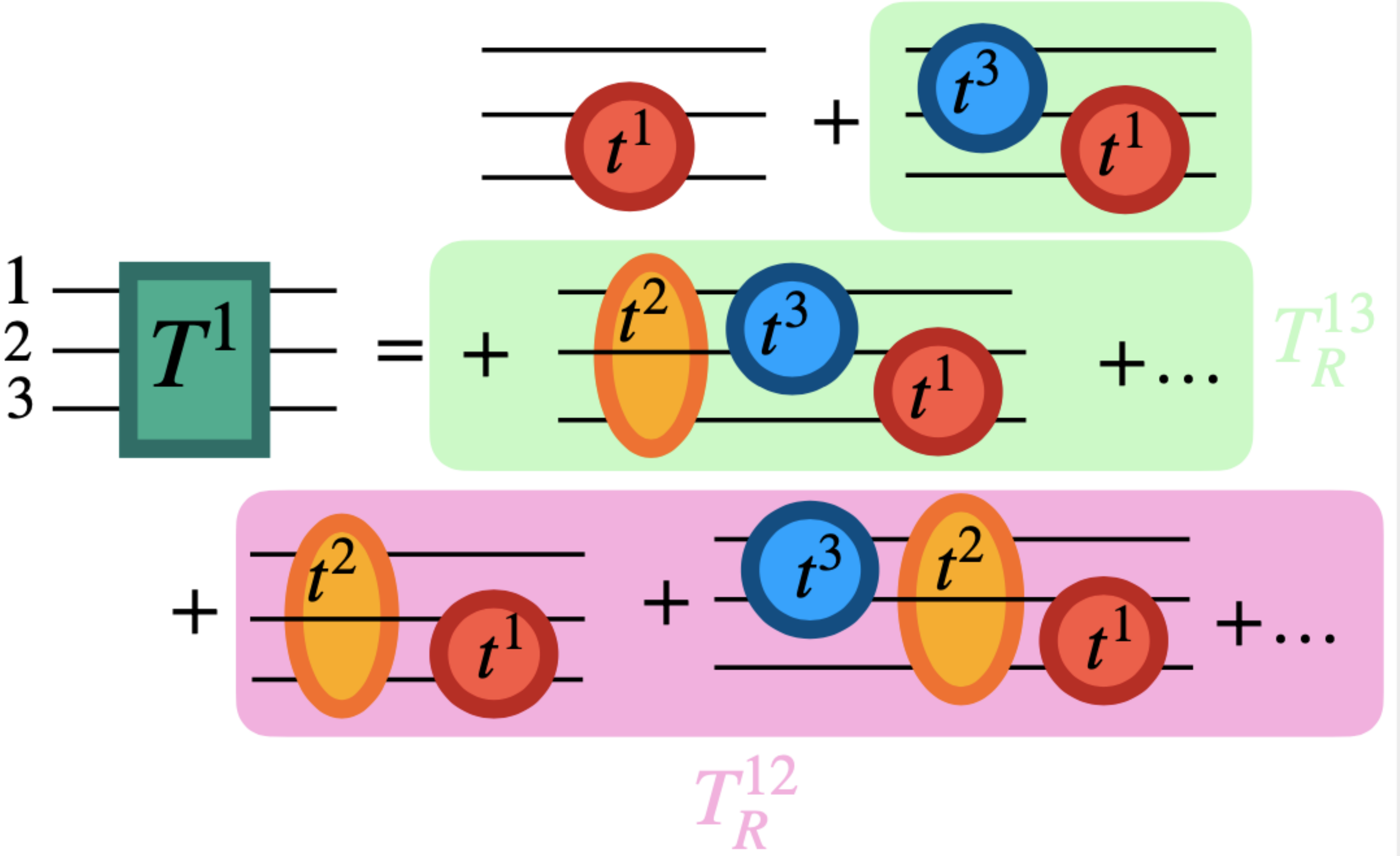}
\caption{Some of the diagrams contributing to the $T^1$ partition. We can distinguish two series involving more than two scattering two-body $t$-matrices, and which are represented by the $T^{12}_R$ and $T^{13}_R$ partitions. These partitions consider all diagrams whose last two right most interactions are given in terms of $t^1$ and $t^2$ ($t^1$ and $t^3$) for $T^{12}_R$ ($T^{13}_R$).}\label{Fad}
\end{figure}

The above mentioned cancellation between the off-shell part of the two-body $t$-matrices and the contact term arising from the Lagrangian can be used to convert the Faddeev equations~\cite{Faddeev:1960su} into a set of six algebraic coupled equations~\cite{alberkan,alber2}, which are given by
\begin{align}
T^i &=t^i\delta^3(\vec{k}^{\,\prime}_i-\vec{k}_i) + \sum_{j\neq i=1}^3T_R^{ij}, \quad i=1,2,3,\label{Ti}\\
T^{\,ij}_R &= t^ig^{ij}t^j+t^i\Big[G^{\,iji\,}T^{\,ji}_R+G^{\,ijk\,}T^{\,jk}_R\Big]. \label{TR}
\end{align}
In Eqs.~(\ref{Ti}), each of the partitions $T^i$ represents an infinite series of contributions to the scattering arising from Feynman diagrams where the \textit{ith} particle is, by convention, a spectator in the right most interaction (see Fig.~\ref{Fad}), such that the three-body $T$-matrix can be obtained as $T=T^1+T^2+T^3$. The $\vec{k}_{i}$ ($\vec{k}^\prime_{i}$) in Eq.~(\ref{Ti}) corresponds to the initial (final) momentum of the particle $i$ and $t^{i}$ is the two-body $t$-matrix which describes the interaction of the $jk$ pair (with $j \neq k\neq i=1,2,3$) and which is obtained by solving Eq.~(\ref{bse}). In Eqs.~(\ref{TR}), $g^{ij}$ represents the three-body Green's function of the system and the $G^{ijk}$ matrix is a loop function involving three hadron propagators as well as a two-body $t$-matrix (see, for example, Ref.~\cite{alberkan} for the particular expressions of $g^{ij}$ and $G^{ijk}$). 

The $T^{ij}_R$ partitions obtained by solving Eq.~(\ref{TR}) are functions of two variables: the center of mass energy of the three-body system, $\sqrt s$, and, for example, the invariant mass of the particles 2 and 3, $\sqrt{s_{23}}$ (since the other two invariant masses, $\sqrt{s_{12}}$ and $\sqrt{s_{31}}$, can be determined in terms of $\sqrt s$ and $\sqrt{s_{23}}$, as shown in Refs. \cite{alberkan,alber2}). To search for a possible formation of a three-body state, we study the behavior of the modulus square of the three-body $T_R$-matrix, $T_R=\sum\limits_{i\neq j=1}^3 T^{ij}_R$, and plot $|T_R|^2$ as a function of $\sqrt{s}$ and $\sqrt{s}_{23}$. A peak appearing in such a plot is interpreted as a state generated from the three-body dynamics involved in the system under investigation.  For convenience, to identify the peaks found with physical states, it is necessary to project first the $T_R$ matrix on an isospin base. In particular, we consider a basis in which the states are labeled in terms of the total isospin $I$ of the three-body system and the isospin of one of the two-body subsystems, for example, the isospin of the subsystem labeled as particles 2 and 3, $I_{23}$, and evaluate the transition amplitude $\langle I,I_{23}|T_R|I,I_{23}\rangle$.

We must stress here that Eq.~(\ref{TR}) is a matrix equation, in general,  and, hence, $T_R$ is a matrix of amplitudes, with its elements corresponding to transitions between different three-body coupled channels.  It is worth mentioning that when studying three-body systems involving mesons, a large number of channels may couple to a given set of quantum numbers. For instance, in Ref.~\cite{alberkan}, to study two-meson--one-baryon  dynamics with total strangeness $-1$ we considered the following coupled channels:  $\pi^0 K^- p$, $\pi^0 \bar K^0 n$, $\pi^0 \pi^0 \Sigma^0$, $\pi^0 \pi^+ \Sigma^-$, $\pi^0 \pi^- \Sigma^+$, $\pi^0 \pi^0 \Lambda$, $\pi^0 \eta \Sigma^0$, $\pi^0 \eta \Lambda$, $\pi^0 K^+ \Xi^-$, $\pi^0 K^0 \Xi^0$, $\pi^+ K^- n$, $\pi^+ \pi^0 \Sigma^-$, $\pi^+ \pi^- \Sigma^0$, $\pi^+ \pi^- \Lambda$, $\pi^+ \eta \Sigma^-$, $\pi^+ K^0 \Xi^-$, $\pi^- \bar K^0 p$, $\pi^- \pi^0 \Sigma^+$, $\pi^- \pi^+ \Sigma^0$, $\pi^- \pi^+ \Lambda$, $\pi^- \eta \Sigma^+$, $\pi^- K^+ \Xi^0$. The motive to include these channels together is to treat effective systems, like, $\pi \Lambda(1405)$, $\sigma \Lambda$, $\sigma \Sigma$,  $\kappa(700) N$, $a_0(980) \Lambda$, $f_0(980) \Lambda$, etc., on the same footing. Treating such systems separately can lead to a poor description of the properties of a three-hadron resonance (or quasi-bound state). Or it can result in an under/over prediction of states since a particular resonance may couple strongly to several of these systems with different two-hadron subsystems resonating in the same (three-body) energy region. 

The situation in three-hadron systems involving mesons, in some sense, is analogous to the studies of two-hadron systems consisting of mesons. Taking examples from the previous sections, we recall that hadrons like $\Lambda(1405)$ and $f_0(980)$ can be understood as those generated from two-hadron coupled-channel dynamics and a good description of their properties requires the consideration of several two hadron coupled channels. For instance, $f_0(980)$ may be described as a $K \bar K$ bound state, but it gets a finite width by including $\pi \pi$ as a coupled channel in the formalism. In the same way, it is important to consider coupled channels when studying three-hadron dynamics.  

 There is, however, a difference in the case of three-hadron studies, since the same state may be seen in different configurations of a given three-body system. For instance, a state found in an effective $K \Lambda(1405)$ channel may also manifest itself in an effective $f_0(980) N$ channel and should not be treated as a different state.  Such findings would in fact imply that both such configurations occur simultaneously. In fact, it is possible that in some cases more than one subsystem acts like a resonance or a bound state. For instance, in Refs.~\cite{kanchan2,jidoalber}, we studied the $N K \bar K$ system and coupled channels and found that a $1/2^+$ nucleon resonance arises with mass 1920 MeV when the two mesons resonate as $a_0(980)$ and the meson-baryon subsystem with strangeness $-1$ is configured as $\Lambda(1405)$. 

One also needs to be careful when restricting interactions of a subsystem in a fixed isospin because one may miss the coupling of the three-body state to other isospin configurations of the subsystem. For example, the same $N^*(1910)$ mentioned above was found to couple to both $f_0(980) N$ and $a_0(980) N$ configurations of the $N K \bar K$ system. 

The success of the consideration of several channels (mentioned above) in Ref.~\cite{alberkan} is clear from the findings of the work, which brought forward a common description for all $1/2^+$ $\Lambda$ and $\Sigma$ states in the 1500-1800 energy region, that is, states arising from three-hadron dynamics.

It should be also useful to mention here that sometimes consideration of several channels may not be necessary. For instance, we studied 14 coupled channels with total strangeness zero in Ref.~\cite{alber2} and found the generation of $N^*(1710)$. However, we found  that considering $\pi \pi N$ interactions alone gave the same results, and  that $N^*(1710)$ can be understood as an effective $\sigma N$ state. Such a description gives a clear explanation for intriguing experimental findings like a large branching fraction of $N^*(1710)$ to the $\pi \pi N$ channel and a smaller decay width to the $\pi N$ channel. Yet another case where we found a three-body channel to give dominant contribution in the formation of a state is $\phi K \bar K$ generating $\phi(2170)$.

\begin{figure}[h!]
\centering
\includegraphics[width=\textwidth]{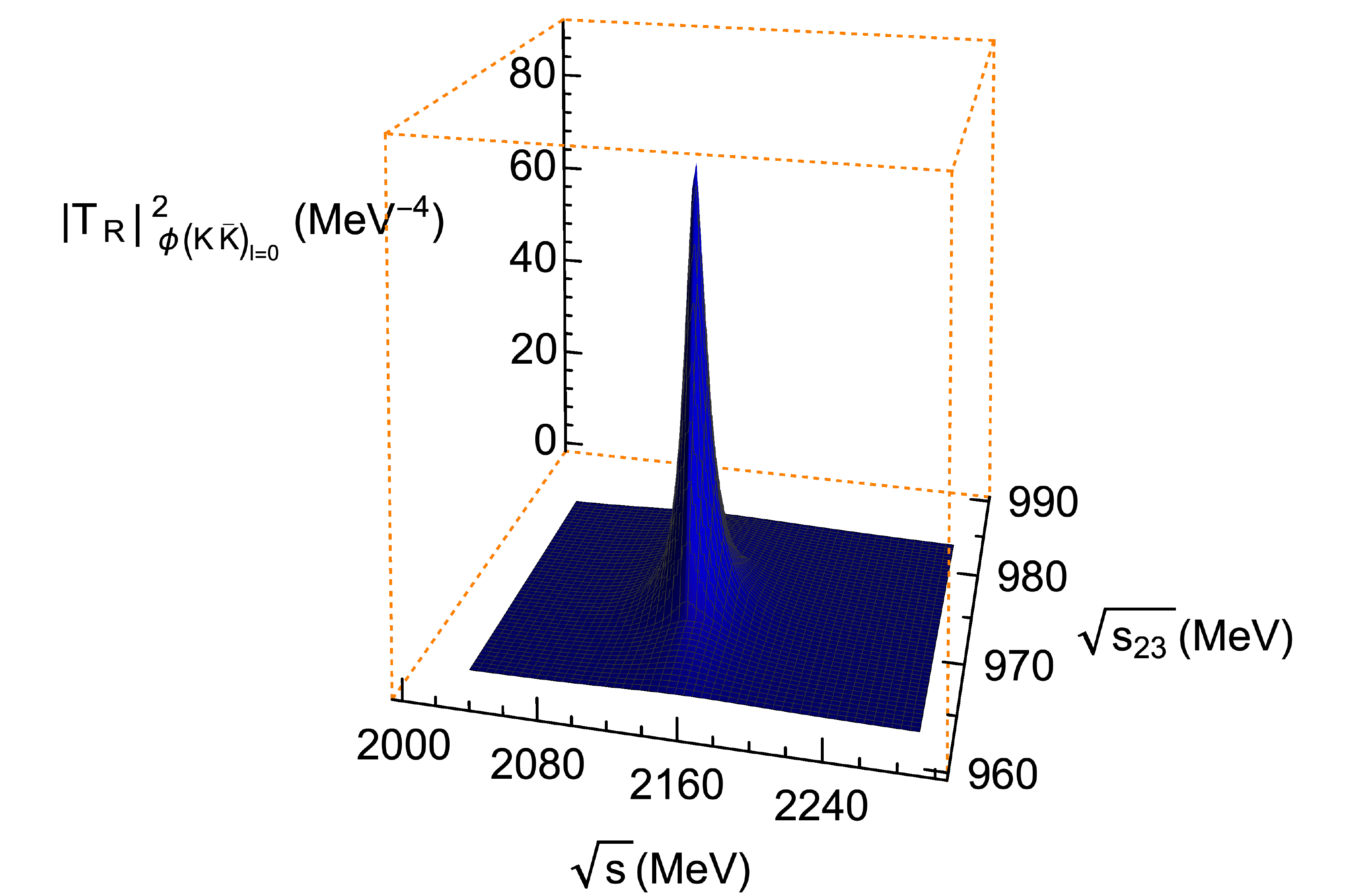}
\caption{$|T_R|^2$ for the $\phi K\bar K$ system. The peak structure seen in the figure can be associated with the generation of $\phi(2170)$ as a $\phi K\bar K$ state, with $K\bar K$ resonating as $f_0(980)$.}\label{phi}
\end{figure}

As an example of the appearance of a three-hadron resonance in the $T_R$ amplitude, we show    $|\langle I=0,I_{23}=0|T_R|I=0,I_{23}=0\rangle|^2$ for the $\phi K\bar K$ system for the configuration $I=0$, $I_{23}=I_{K\bar K}=0$ in Fig.~\ref{phi}. As can be seen, a peak for $\sqrt{s}\simeq 2150$ MeV is observed when the invariant mass of the $K\bar K$ subsystem in isospin 0 is $\sqrt{s_{23}}\simeq 970$ MeV. Such result shows the generation of a three-body $\phi K\bar K$ state when the $K\bar K$ subsystem in isospin 0 forms the $f_0(980)$. The properties of this state, as shown in Ref.~\cite{phi2175}, are compatible with the $\phi(2170)$ listed by the Particle Data Group~\cite{pdg}. 

Within this approach several three-hadron systems, like $D D K$, $J/\psi K\bar K$, $\pi K\bar K$, have been studied and formation of states which can be cataloged as molecular three-hadron bound states or resonances have been found (see Table~\ref{table} for the different systems studied). It is important to emphasize that in this formalism only the on-shell part of the two-body $t$-matrices is necessary to solve Eqs.~(\ref{TR}). Such two-body $t$-matrices do not have to be obtained necessarily from a theoretical model, and one could decide to solve Eqs.~(\ref{TR}) by using two-body $t$-matrices determined directly from experimental phase-shifts and inelasticities, if necessary, as shown in Ref.~\cite{kanchan2}.

\section{Fixed center approximation and many meson systems}

Certainly, the most accurate and formally well based technique to tackle the three-body interaction is the Faddeev equations reviewed in the previous sections. However, in many instances it is not practical or simple to implement it, either analytical or numerically. Nevertheless, in some systems of three strongly interacting particles, it may  be the case that two of those particles are more strongly correlated among themselves (or quasibound) than with the third one. In such systems, with a strong bound pair plus a loosely bound third particle, a much simpler approximation can be applied which goes by the name of Fixed Center approximation (FC). Even though the theory  for the rescattering of one particle on a pair of fixed heavier ones has been known 
since as earlier as 1945 \cite{Foldy:1945zz} and applied to pion deuteron scattering few years later
\cite{Brueckner:1953zz,Brueckner:1953zza}, and to the $K^-$ deuteron system
\cite{Chand:1962ec},
it has not attracted much preaching or predilection among the many body community. It has not been since twenty years ago, when it  has been revitalized because of the interest in its application to the kaonic deuterium \cite{Barrett:1999cw,Deloff:1999gc}. A special boost in the wide implementation of the FC, was given in Ref.~\cite{Kamalov:2000iy} with the merger of the FC techniques and the use of chiral unitary amplitudes for the $\bar K N$ interaction. Since then, a plethora of works implementing several meson and/or baryon systems using for the two-body scattering the  amplitudes obtained from the chiral unitary approach have flourished, mainly focused on studying the possible dynamical generation of three-body resonant states, even with incursions into the charm and beauty sectors, (see the works labeled as FC in Table~\ref{table}). 

Since it is not the aim of the present work to provide an exhaustive review of all these approaches but to motivate the use of this technique for interested people in the topic, we will illustrate the main features of the FC with an example: systems made of several (up to six) $\rho(770)$ mesons, extracted from Ref.~\cite{luismulti}. On the way, we will address the question raised in the Introduction regarding whether it is possible  to have a nucleus-like system of mesons. The details of the formalism can be found in Refs.~\cite{luismulti,Kudryavtsev:2016pzj}. This system is particularly simple and illustrative of the idea underlying the FC and it allows one to extend the formalism iteratively to more than three particles. 
Although the FC has been typically applied to systems where the third particle is lighter than those forming the bound state (cluster), it was shown in Ref.~\cite{MartinezTorres:2010ax} that 
the approach is effective when there is not enough energy to excite
the cluster, {\it i.e.} the cluster remains bound. In the particular case of the $\rho\rho$  system,  it was found in Ref.~\cite{raquel} that two $\rho$ mesons bind very tightly for spin $S=2$ forming the $f_2(1270)$ resonance. This implies a very strong binding energy per $\rho$ meson, of about 135~MeV, and makes it advisable to apply the FC approximation to the three-$\rho$ system.

 \begin{figure}
 \centering
 \includegraphics[width=0.99\textwidth]{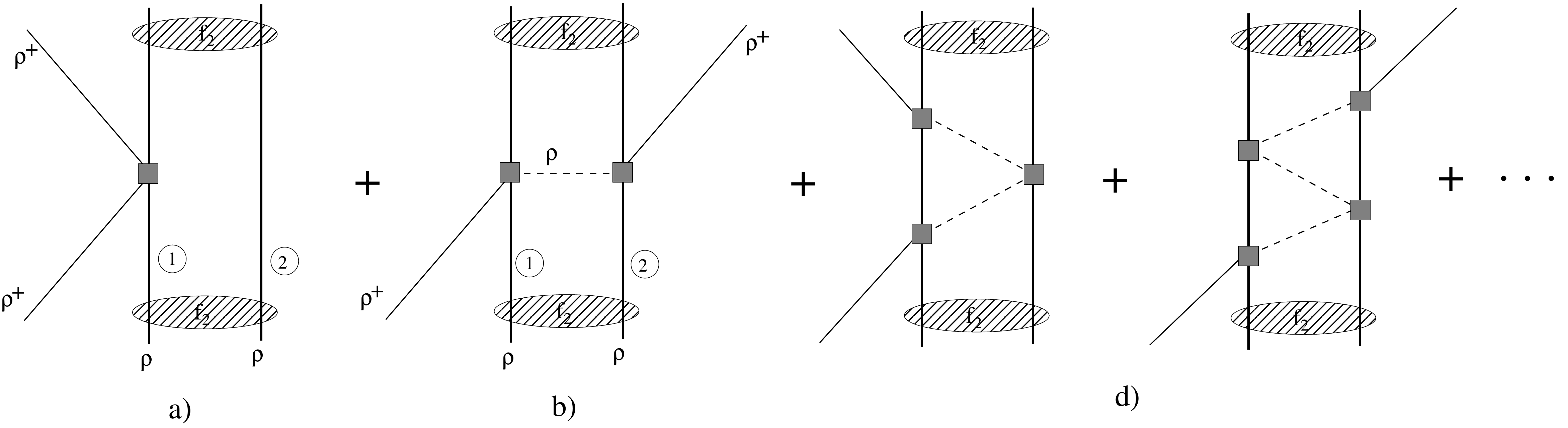}
 \caption{ Diagrammatic representation of the FC approximation. Diagrams
a) and b) represent the single and double scattering contributions respectively.
}\label{FCA_diag}
 \end{figure}

A diagrammatic representation of the mechanisms implied in the FC is shown in Fig.~\ref{FCA_diag}. An external $\rho$ meson reinteracts repeatedly with each of the other two which make up the 
$f_2(1270)$ cluster. In this situation, the Faddeev equations (\ref{Ti}) and  (\ref{TR}) are approximated by 
\begin{align}
T_1&= t_1+t_1 G_0 T_2 \nn \\
T_2&= t_2+t_2 G_0 T_1  \nn \\
T&=T_1+T_2  \label{eq:FCA}
\end{align}
which are written in terms of two partition functions $T_i$ which account for all the diagrams starting with the interaction
of the external $\rho$-particle
 with particle $i$ of the compound $f_2(1270)$
system  and which sum up to the total amplitude\footnote[4]{Alternative equations are obtained in Ref.~\cite{Sekihara:2016vyd} by considering more partitions, depending on whether the final particle is emitted on the left or the right of the cluster in the diagrams.} $T$. The two-body amplitudes $t_i$ represent the $\rho\rho$ unitarized scattering 
amplitudes from \cite{raquel}.
The diagram in Fig.~\ref{FCA_diag}a represents the
single-scattering mechanism 
($t_1$ in Eq.~(\ref{eq:FCA})), which might remind some readers of the impulse approximation in meson-deuteron scattering,
and Fig.~\ref{FCA_diag}b is
the double-scattering mechanism (the next contribution, $t_1=t_1+t_1G_0t_2$). 
The total partition function $T_1$ is finally obtained  by adding the full resummation
of mechanisms of Fig.~\ref{FCA_diag}c. 
In  Ref.~\cite{luismulti} it was shown that $G_0$ can be written as
\ba
G_0\equiv\frac{1}{M_{f_2}}\int\frac{d^3q}{(2\pi)^3} 
F_{f_2}(q)
\frac{1}{{q^0}^2-\vec{q}\,^2-m_\rho^2+i\epsilon}.
\label{eq:G0}
\ea
where $F_{f_2}(q)$ is the $f_2(1270)$ form factor deduced from a potential of the kind of  Eq.~\eqref{eq:Vcutoff} \cite{luismulti}.

Eqs.~\eqref{eq:FCA} and \eqref{eq:G0} are meant to provide directly the three-body amplitudes. There is, however, a caveat in its derivation. If we look at the third diagram of Fig.~\ref{FCA_diag}, Eqs.~\eqref{eq:FCA} imply that the two-body amplitude of the second square from down up is the same as the one of the first square. This is not necessarily true, although one hopes it to be a good approximation. This assumption is good for the second diagram because in both scatterings one line is external and the other one internal. So, the problem stems from the third diagram on and one can see the problem by going explicitly to the three scattering term assuming the lines of the cluster to be always in the ground state of the cluster. By doing that, one can see that the approximation is very good if the range of the interaction is smaller than the separation of the particles of the cluster. Another way to face it is to see that the assumption is correct if the scattering amplitude with the bound particle can be written in a separable way $F(q)F(q')$ for the second square of the third diagram ($q$ , $q'$, entering and outgoing exchanged lines), something that can be justified in cases where one studies bound sates (see Ref.~\cite{Xie:2016zhs}). This is interesting to mention because in other schemes, like the one used in \cite{Kamalov:2000iy} or \cite{Baru:2012iv} the partitions $T_{1,2}$ in Eqs.~\eqref{eq:FCA} are treated as operators which are sandwiched with the cluster wave function at the end. In these schemes the cluster wave function is only meant at the beginning and the end of the diagrams in Fig.~\ref{FCA_diag} and in the intermediate states the cluster is allowed to be excited. Our approach is not new and would be equivalent to other variants of the FC used in the literature, like the FCA-average \cite{Deloff:1999gc}. In cases when one has a relatively large binding it is unlikely that the cluster is excited in intermediate states, in which case one should expect our approach to be very good. For example, systems like $N K \bar K$ and $D K \bar K$ have been studied both within the FCA approach and by solving the Faddeev equations and similar results have been found in both cases \cite{kanchan2,jidoalber,xiealber,albermarina,Debastiani:2017vhv}. It is also clear that if the dynamics involved in the three-body system produces a state well above the threshold, the method might be unreliable, as it was found explicitly in Ref.~\cite{MartinezTorres:2010ax}. 

\begin{figure}[h]
\begin{center}
\subfigure[
Masses of the states obtained \cite{luismulti} with the FC as a function of the
number of constituent $\rho(770)$ mesons, $n_\rho$. Only single
scattering contribution (dotted line); full model (solid line);
experimental values from the PDG~\cite{pdg}, (circles).]{\includegraphics[width=.45\textwidth]{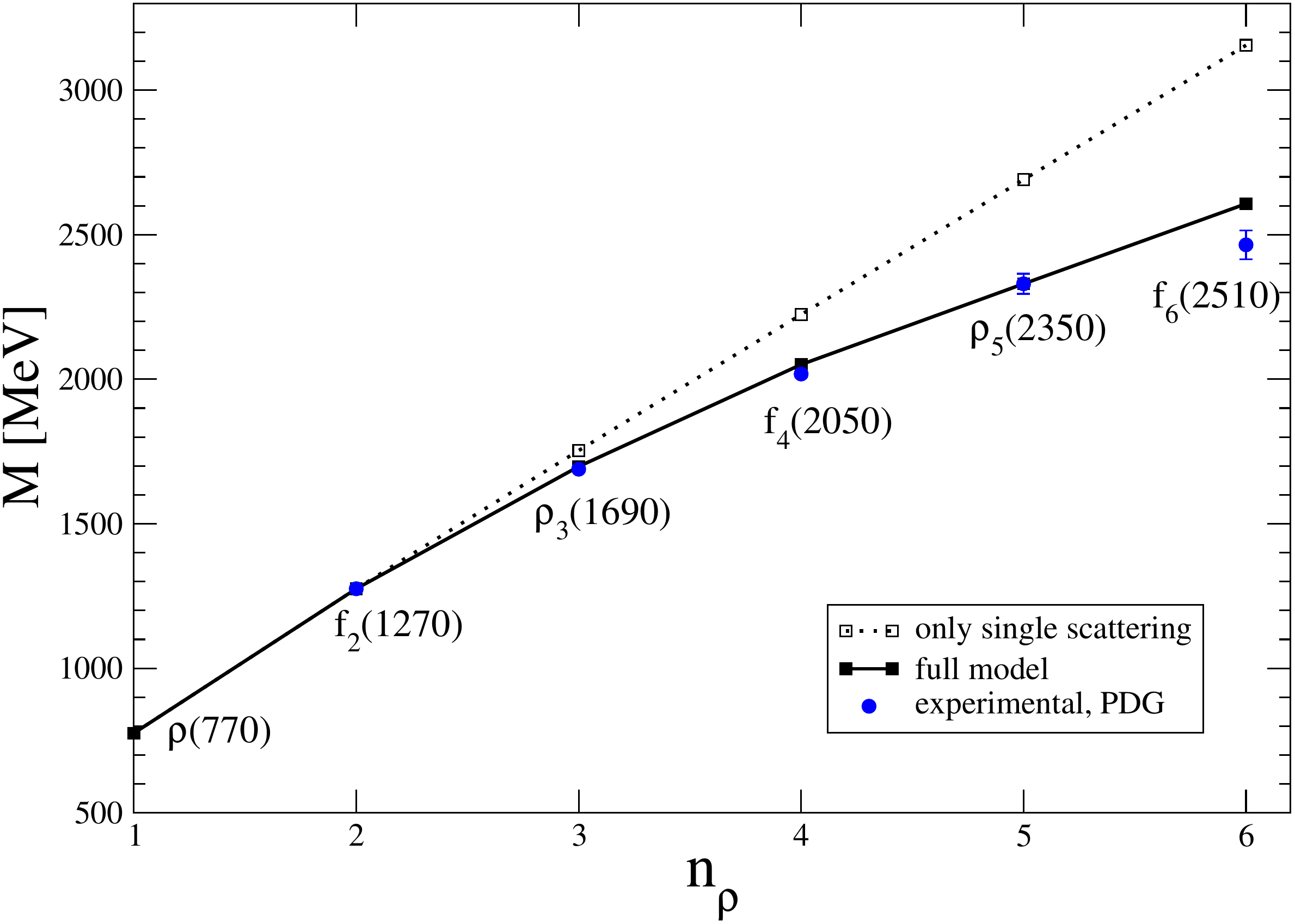}}
\hspace{3mm}
\subfigure[Modulus squared of the unitarized $K^*$--multi-$\rho$ amplitudes leading to the $K^*_5(2380)$]{\includegraphics[width=.48\textwidth]{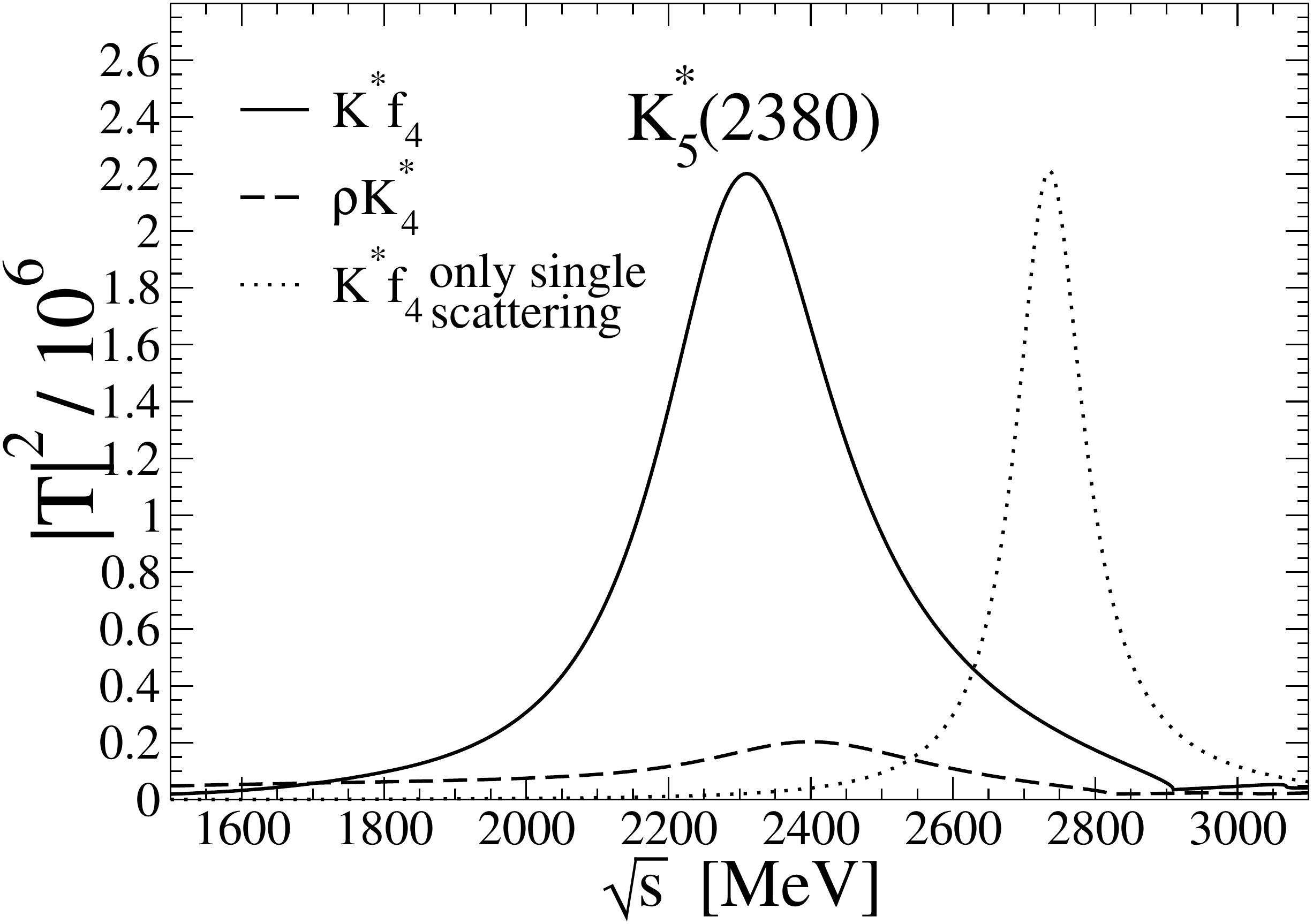}  }
\caption{Some example results for the Fixed Center approximation.}
\end{center}
\label{fig:FCAres}
\end{figure}

Coming back to the problem of the multirho systems,
with three $\rho$ mesons interacting in the way described above, a resonant shape for $|T|^2$  peaking at 1698~MeV was obtained in Ref.~\cite{luismulti} which can be identified with the experimental $\rho_3(1690)$ meson. 
One may wonder whether one can get further quasibound systems with more $\rho$ mesons. The next step would be to study four $\rho$ meson systems which, given the strong tendency of two $\rho$ mesons with parallel spins to clusterize into an $f_2(1270)$, the natural approach would be to consider the interaction of two clusters of two $\rho$ mesons, each of them forming a $f_2(1270)$. This is also done in  Ref.~\cite{luismulti} by iterating the process in Fig.~\ref{FCA_diag}, where now the external $\rho$ is substituted by another $f_2(1270)$ cluster of two $\rho$ mesons. Since again a bound system is obtained, we can iterate the process and proceed analogously adding more $\rho$ mesons. A plot of the resonant masses obtained for systems of increasing number of $\rho$ mesons compared to experimental values for corresponding known resonances is shown in Fig.~\ref{fig:FCAres}a.
Note that the multi-step mechanisms (Figs.~\ref{FCA_diag}b and c) are crucial in order to obtain a remarkable agreement with experimental values. We can also realize that, as the number of constituents increases, the generated states have an increasingly larger binding energy per $\rho$ particle, which points out at a possible saturation for some number of mesons for which it would be energetically free to add another $\rho$ meson, like in a condensate. However the large width of the $\rho$ meson and of the generated multi-$\rho$ states could make the hypothetical condensate to fade away. Actually what we see in practice is that the widths of the states increase with the number of $\rho$ mesons and the width of the $f_6$ state is already so large, around 500~MeV, that its identification as a particle is already at its limits.

On the other hand, the interaction of a $K^*(892)$ and a $\rho$ meson in spin $S=2$ also binds forming the $K^*_2(1430)$ \cite{gengvec}. In an analogous way to the multi-$\rho$ meson system explained above, in Ref.~\cite{kstarmulti} the interaction of a $K^*(892)$ with an increasing number of $\rho$ mesons was studied. 
Now, since we have two very strongly attractive two-body structures, $f_2$ and $K^*_2$, for the three-body interaction in FC we have the possibilities of the interaction of a  $K^*$ with the $f_2$ (made of two $\rho$'s) or the interaction of a $\rho$ with a $K^*_2$. As another example, for the five body interaction we can interact the $K^*$ with an $f_4$ (bound state of two $f_2$ which comes up from the four $\rho$ interaction) or a $\rho$ with a $K^*_4$ (made of an $f_2$ and a $K^*_2$ bound system in the four body amplitude). We show, as an example, in Fig.~\ref{fig:FCAres}b the amplitudes for these two possibilities and how a resonant structure for the  $K^*_5(2380)$ arises. We also see the important effect of the mechanisms beyond the single step process. 
Note that, in principle,  those different possible channels, like in the case of the $K^*_5$, should be considered coupled into the equations. But that would require to break the $K^*_4$ and the $f_4$ which would invalidate the FC assumptions of not altering much the clusters. That would mean to solve the full Faddeev equations we were trying to avoid, or of one more involved quantum field theory calculation. However if the  amplitude  of one of the channels is much smaller than the dominant one (as is the case of the example shown in Fig.~\ref{fig:FCAres}b) or it produces the peak at a similar energy, it was shown in Ref.~\cite{Roca:2011br} that the FC is still a good approximation.
Interacting up to six vector mesons,  resonant structures where obtained \cite{kstarmulti} which could be associated to the experimental $K^*_2(1430)$, $K^*_3(1780)$, $K^*_4(2045)$, $K^*_5(2380)$ in addition to the prediction of an 
undiscovered $K^*_6$ with quantum numbers $I(J^P)=1/2(6^+)$, a mass between $1650-2750$~MeV and a width around several hundreds MeV. As in the multi-$\rho$ system, the width of the states increases with the number of vector mesons added to the ensemble,  making it increasingly more difficult to identify experimentally.

\section{Compilation of few-body systems studies involving mesons}
    As we said in the Introduction, this paper is not a review paper, but one showing how to make advantage of useful tools to navigate in the world of few-body systems involving several mesons and systems in the heavy quark sector. However, we present a table just to show which works have been studied in this field to facilitate the work of few-body practitioner wishing to join efforts in this field. 
\begin{longtable}{llcl}
\caption{Few-body systems studied in the literature involving one, two or more mesons. $P$: Pseudoscalar; FC: Fixed Center; $\chi F$: Chiral Faddeev; V: Variational; GE: Gaussian expansion; QSR: QCD sum rules; BO: Born-Oppenheimer}\label{table}\\
Components& States generated& Method used& References\\
\hline\hline
$\bar KNN$& $\bar K$ bound states& $\begin{array}{l}\text{F}\\\text{V}\\\text{FC}\end{array}$&$\begin{array}{l}\text{\cite{sato,mares,suslov,revai}}\\\text{\cite{hyodo,ikeda,kamano,inoue,doteinoue}}\\\text{\cite{bayar,bayar2,bayar3,bayar4}}\end{array}$\\\\
$2PN$&$\begin{array}{l}1/2^+~\Sigma,~\Lambda~\text{excited}\\1/2^+~N^*~\text{states}\\(N^*(1920))\end{array}$& $\begin{array}{l}\text{$\chi$F}\\\text{$\chi$F}\end{array}$&$\begin{array}{l}\text{\cite{alberkan}}\\\text{\cite{kanchan2}}\end{array}$\\\\
$\pi\pi N$&$N^*(1710)$&$\chi$F&\cite{alber2}\\\\
$K\bar K N$&$N^*(1920)$&$\begin{array}{c}\text{V},\\\text{FC}\end{array}$&$\begin{array}{l}\text{\cite{jido,jidoalber}}\\\text{\cite{xiealber}}\end{array}$\\\\
$KK\bar K$&$K(1460)$&$\chi$F&\cite{alberjido}\\\\
$\pi K\bar K$, $\pi\pi\eta$&$\pi(1300)$, $f_0(1790)$&$\chi$F&\cite{kanchanjido}\\\\
$\phi K\bar K$, $\phi\pi\pi$&$\phi(2170)$&$\chi$F&\cite{phi2175}\\\\
$\pi\rho\Delta$&$\Delta_{5/2^+}(2000)$&FC&\cite{xiepedro}\\\\
$\pi\bar K K^*$&$\pi_1(1600)$& FC&\cite{xiexurong}\\\\
$\eta\bar K K^*$&$\begin{array}{l}0(1^-)~\text{state around}\\\text{1700 MeV}\end{array}$&FC&\cite{zhangxie}\\\\
$\rho K\bar K$&$\rho(1700)$&FC&\cite{bayarliang}\\\\
multi-$\rho$&$\begin{array}{l}f_2(1270),~\rho_3(1690),~f_4(2050),\\\rho_5(2350),~f_6(2510)\end{array}$&FC&\cite{luismulti}\\\\
$K^*$ multi-$\rho$&$\begin{array}{l}K^*_2(1430),~K^*_3(1780)\\K^*_4(2045),~K^*_5(2380),~K^*_6\end{array}$& FC&\cite{kstarmulti}\\\\
$PVV$&$\begin{array}{l}	\pi_2(1670),~\eta_2(1645),~K^*_2(1770)\end{array}$& FC&\cite{Roca:2011br}\\\\
$K$ multi-$\rho$&several $K^*$ states&FC&\cite{kmulti}\\\\
$DNN$&$D$ bound state&$\begin{array}{l}\text{FC}\\\text{V}\end{array}$&$\begin{array}{l}\text{\cite{xiaoka}}\\\text{\cite{xiaobayar}}\end{array}$\\\\
$\begin{array}{l}NDK,~ND\bar K,\\N D\bar D\end{array}$&$\begin{array}{l} \text{bound states of}\\\text{3050, 3150, 4400 MeV}\end{array}$& FC&\cite{xiaobayar}\\\\
$\begin{array}{l}DDK,~DD_s\eta,\\DD_s\pi\end{array}$& $\begin{array}{l}I=1/2~\text{state}\\\text{around 4140 MeV}\end{array}$& $\chi$F&\cite{gengalber,liualber}\\\\
$DDK$& Bound state, $B\simeq 70$ MeV& GE&\cite{pavon}\\\\
$DDDK$& Bound state, $B\simeq 180$ MeV&GE&\cite{pavon}\\\\
$J/\psi K\bar K$&$Y(4260)$&$\chi$F&\cite{alberdani}\\\\
$KD\bar D^*$&$K^*$ bound states& FC&\cite{renkanchan}\\\\
$DK\bar K$&$D$-like state at 2900 MeV&$\begin{array}{c}\text{QSR,}~\chi\text{F}\\\text{FC}\end{array}$&$\begin{array}{c}\text{\cite{albermarina}}\\\\\text{\cite{Debastiani:2017vhv}}\end{array}$\\\\
$\rho D\bar D$&$I=0,\,1$ states 4200-4300 MeV& FC&\cite{durkaya}\\\\
$\rho B^*\bar B^*$&$J=3$ state at 10950 MeV& FC&\cite{bayarpedro}\\\\
$D^{(*)}B^{(*)}\bar B^{(*)}$&Several bound states&FC&\cite{luiscola}\\\\
$BD\bar D$, $BDD$&$BD\bar D$ bound state $\sim$ 8950 MeV&FC&\cite{diasroca}\\\\
$\begin{array}{l}BB^*B^*,\\B^*B^*B^*\end{array}$&Bound $C=3$ meson&F&\cite{valcarce}\\\\
$DD^*K$, $BB^*\bar K$&$\begin{array}{l}\text{Bound states}\\\text{4318 MeV, 11014 MeV}\end{array}$&BO&\cite{ulfdeloca}\\\\
$\bar K^*B\bar B$, $\bar K^*B^*\bar B^*$&Several bound states&FC&\cite{Ren:2018qhr}\\\\
$BBB^*$&probable bound state&BO&\cite{delocaplus}\\\\
$D$ multi-$\rho$& seven $D^*$ states& FC&\cite{xiao}
\end{longtable}
\section{Conclusions}
 In the present work we have not attempted to make a review of recent many body works in hadron physics. The purpose has been different and the work is meant to establish a bridge between few-body practitioners working in traditional areas of few-body physics and, on the other hand, people working on problems related to recent findings of exotic hadrons in different laboratories around the world. We have started by giving a pedagogical introduction to show how the interaction of pairs of particles, which is the starting point in a many body problem, can be determined using tools of chiral unitary theory and its extensions to the heavy quark systems. We have discussed how chiral dynamics is suited to show cancellations between three-body contact terms, provided by the theory, and terms stemming from the contribution of the off-shell part of the two-body amplitudes in the Faddeev equations. This finding is most welcome from two different perspectives: first, it removes ambiguities related to the off-shell parts of the amplitudes, which are unphysical, and hence the final results cannot depend upon them. Second, it leads to an enormous simplification by allowing the use of only the on-shell parts of the two-body amplitudes as inputs in the few-body calculations. 

     Further, we have shown two useful methods to deal with such few-body systems. One consists of solving the Faddeev equations, which are suited to using, as input, two-body amplitudes obtained within the chiral unitary approach in coupled channels or its extension to the heavy quark sector. The other one is solving the Faddeev equation by considering the Fixed Center Approximation, which leads to a much simpler formalism and is convenient in cases of bound few-body systems, and is certainly recommendable in a preliminary research of unexplored systems. 

      We also provide a short discussion on the multi-meson systems, which is an emerging field, with two considerations in mind: the meson-meson interaction is in some cases much stronger than the nucleon-nucleon interaction: unlike the ordinary nuclei, where the baryon conservation number is a stabilizing factor, there is no meson number conservation. Thus, multi-meson systems can decay to fewer mesons for which a phase space is always available. The only thing that prevents more and more mesons to stick together is the decay width of such systems. The issue then becomes how many mesons can we put together until the width is so large that its identification as a resonance becomes impractical. 

  Finally, we have made a list of recent works on unconventional few-body systems involving mesons and the few-body techniques used to study them, with the purpose to guide researchers into the study of new systems, or systems already studied but with a different technique. There is no doubt that the impressive advance made in hadron physics in recent years due to the experimental discoveries in modern facilities, will have its impact on the study of new few-body systems. With this work we wish to motivate hadron physicists, mostly working on two-body systems, to step towards studying few-body systems too, and to traditional few-body physicists, with their excellent few-body tools, to venture in the new field of more exotic systems for which we envisage a bright future.

\bibliographystyle{unsrt}
\bibliography{ref.bib}

\end{document}